# Atomically Resolved Observation of Continuous Interfaces between As-grown MoS$_2$ Monolayer and WS$_2$/MoS$_2$ Heterobilayer on SiO$_2$

Fan Zhang[1†], Zhixing Lu[2†], Yichul Choi[1], Haining Liu[3], Husong Zheng[1], Liming Xie[3], Kyungwha Park[1], Liying Jiao[2*], Chenggang Tao[1*]

[1]Department of Physics, Virginia Tech, Blacksburg, Virginia 24061, USA

[2]Department of Chemistry, Tsinghua University, Beijing 100084, China

[3]Key Laboratory of Standardization and Measurement for Nanotechnology of Chinese Academy of Sciences, National Center for Nanoscience and Technology, Beijing 100190, China

[†]These authors contributed equally to this work.

[*]Corresponding to: lyjiao@mail.tsinghua.edu.cn and cgtao@vt.edu.

**Abstract**

Van der Waals (vdW) heterostructures synthesized through the chemical vapor deposition (CVD) method allow creation and tuning of intriguing electronic and optical properties of two-dimensional (2D) materials. Especially, local structures in the heterostructures, such as interfaces, edges and point defects, are critical for their wide range of potential application. However, up to now atomic scale measurements of local structures in as-grown 2D heterostructures on insulating substrates are still rare. Here we report our scanning tunneling microscopy (STM) and spectroscopy (STS) study of as-grown MoS$_2$ monolayer and WS$_2$/MoS$_2$ heterobilayer on SiO$_2$. The heterobilayer appears smoother than the MoS$_2$ monolayer, with root mean square (RMS) roughness of 0.230 nm in the former and 0.329 nm in the latter. For the first time to our knowledge, we directly observed a novel type of continuous interfaces between the MoS$_2$ monolayer and the top layer of the heterobilayer with atomic resolution. Our STS results and density functional theory (DFT) calculations revealed the band gaps of the heterobilayer and the MoS$_2$ monolayer. The finding of the continuous interfaces and the systematic characterizations could have significant impacts on optimizing and designing new 2D heterostructures.

**Keywords:** 2D heterostructures, transition metal dichalcogenides, interface, roughness, scanning tunneling microscopy, and scanning tunneling spectroscopy.

**Introduction**

Two-dimensional (2D) transition metal dichalcogenides (TMDs), ranging from metals to semiconductors, have attracted a huge amount of research efforts due to their unique properties e.g., charge density waves, direct band gaps, and valleytronics[1-3]. These properties are critical for building electronic, spintronic and optical devices, such as single photoemission devices, monolayer TMD transistors and photodetectors[4-11]. Beyond structures composed of single chemical component, heterostructures made of distinct layers bound by van der Waals (vdW) interlayer interaction define a new category of 2D materials, which often host novel physical properties not found in the constituent parts. In practice, designed vdW heterostructures with desirable properties will broadly extend novel applications of 2D materials and constitute central building blocks of future electronic and optoelectronic devices[12]. In the past years, vdW heterostructures, such as MoS$_2$/graphene and WS$_2$/MoS$_2$, have been intensively studied[12-14]. Among the synthesized vdW heterostructures, WS$_2$/MoS$_2$ heterobilayer (Fig. 1a), particularly demonstrates interesting optical and electronic behaviors such as ultrafast interlayer photon transfer[14], and can be used as heterojunction transistors and solar cells[15-17].

VdW heterostructures can be composed through mechanical stacking of individual single-component layers[12, 18]. However, mechanical stacking suffers from limited sizes and randomly generated locations of individual exfoliated flakes. Such a preparation procedure is hence neither scalable nor controllable, and is impractical for applications. This problem finds a solution in the chemical vapor deposition (CVD), a recently developed method for synthesizing vdW heterostructures[19-21]. For WS$_2$/MoS$_2$ heterostructures, both vertical and lateral configurations have been synthesized through CVD[14, 20-23]. In the vertical WS$_2$/MoS$_2$ heterostructres, it was believed that the WS$_2$ layer lies on top of the MoS$_2$ layer with open edges[20]. However, direct evidence for such interfacial structures at the atomic scale has yet been missing, mainly due to technical challenges. For example, transmission electron microscopy[6], another major probing technology with atomic resolution, cannot image as-grown samples directly without transferring. However, the transferring process usually alters certain properties of the as-grown samples, such as roughness and strain induced by the growth substrate.

In this work, we used scanning tunneling microscopy/spectroscopy (STM/STS) combined with an optical localization method to investigate the interfaces between the as-grown $MoS_2$ monolayer and $WS_2/MoS_2$ heterobilayer on $SiO_2$ prepared via CVD. Roughness of the heterobilayer was determined to be $0.230 \pm 0.021$ nm, compared with a roughness of $0.329 \pm 0.033$ nm in the $MoS_2$ monolayer. At the interface between the $MoS_2$ monolayer and the heterobilayer, we observed continuation of the top layer of the heterobilayer into the $MoS_2$ monolayer. To our best knowledge, it was the first atomically resolved observation of such a closed, rather than open, interface between the monolayer and the top layer of the heterobilayer, indicating that the two layers probably connected via covalent bonds. This observation contradicts the previous description of the interface, i.e., the top layer (e.g., $WS_2$) staying on the bottom layer (e.g., $MoS_2$) with an open edge. This finding is important for understanding the growth mechanism of 2D heteromaterials and could potentially explain some transport and optical properties of these materials. We further characterized the electronic properties of the $MoS_2$ monolayer, the heterbilayer, as well as the continuous interface between them. Our results could be crucial for device-related applications involving CVD-synthesized 2D materials and vdW heterostructures.

**Results and discussions**

The $WS_2/MoS_2$ heterobilayer was synthesized via CVD[24]. The optical and atomic force microscopy (AFM) images showed typical $WS_2/MoS_2$ heterobilayer islands surrounded by $MoS_2$ monolayers (Fig. 1b). The heterobilayer islands primarily have triangular shape, indicating a S-rich condition in the growth processes[25]. The $WS_2/MoS_2$ heterobilayer typically was assumed with a twisted angle of 0° or 60°, corresponding to AA' and AB stacking, respectively. To confirm the components of the monolayer and bilayer regions, we performed Raman (Figs. 1c-e) and photoluminescence[2] measurements (Fig S3). In the Raman spectrum (Fig. 1c), the high-frequency peaks are derived from the $MoS_2$ and $WS_2$ layers in the heterobilayer. The four low-frequency peaks corresponded to the shear modes (17.25 cm$^{-1}$ and 22.97 cm$^{-1}$) and the breathing modes (28.68 cm$^{-1}$ and 37.66 cm$^{-1}$) caused by the interlayer interactions within the heterobilayer, as illustrated in Fig. 1d. The Raman and PL mapping (Figs. 1e, S3) images clearly distinguished regions of MoS$_2$ monolayer from those of WS$_2$/MoS$_2$ heterobilayer. Since the heterobilayer was formed by the sequential growth of MoS$_2$ and WS$_2$[15], the WS$_2$ layer is expected to be on the top of the MoS$_2$ layer. To confirm the stacking order of these two layers, we etched the heterobilayers with oxygen at a temperature of 320 °C and identified the compositions of the two layers based on their distinctive oxidation behaviors (Fig. 1f). Although both MoS$_2$ and WS$_2$ can be oxidized at this temperature, the oxidation products of MoO$_x$ were vaporized while the WO$_x$ remained on the samples as a result of their dramatically different sublimation temperature[15]. Since triangular etching pits were only observed in the single layer region (Fig. 1f), we can conclude that the bottom layer is MoS$_2$. In addition, after etching, the Raman intensity for MoS$_2$ dramatically decreased in the single layer region, while it remained the same in the bilayer region (Fig. S4), which further confirms that WS$_2$ is on the top of MoS$_2$ in the bilayer region.

Next, we used STM/STS to characterize the samples at atomic resolution. In our experiments, electrical contacts were thermally deposited through a shadow mask (Fig. 2a). A long focus optical microscope (Infinity K2) guided the STM tip onto individual heterobilayer islands, allowing the edges of the heterobilayer islands to be precisely located in large-scale STM survey (Fig. S5). Fig. 2b shows an STM image of a region containing both heterobilayer and MoS$_2$ monolayer. When zooming into the interfaces between the heterobilayer and MoS$_2$ monolayer, we observed, strikingly, a closed or continuous edge configuration (Fig. 2c). The atomically resolved STM image clearly exhibited that the top layer on the heterobilayer is seamlessly connected to the adjoining MoS$_2$ monolayer, implying a closed edge between the two regions (as illustrated in Fig. 2d). The continuity of the interfaces was observed on each edge of heterobilayer islands (Fig. S6b). The topography of these continuous interfaces between the monolayer and heterobilayer differs dramatically from the open edges of TMDs reported in the previously studies[26-28], suggesting that the two layers were connected by covalent bonds at the edges of the top layer during the high temperature growth. The formation of new chemical bonds at the edges of 2D TMDs has been reported previously and was utilized to synthesize in-plane 2D heterostructures through a two-step CVD growth[29-32]. The majority of the interfaces laid along the zigzag orientation, indicating that the zigzag edges were energetically more favorable to form during the growth. This corroborates the previous theoretical calculations that the zigzag edge has a lower energy than the armchair edge[33]. The interfaces were ~ 4 nm wide, and meandering slightly. The width and meandering could result from the underlying $SiO_2$ substrate being rough and/or the edges of the bottom layer being not atomically sharp. In addition to the interfaces between the heterobilayer and the monolayer, we also observed narrow trenches demarcating regions of bilayers (Figs. 2e, 2f and S6d). The trenches mostly laid along the zigzag orientation and were occasionally observed along the armchair orientation (Fig. S6d). These trenches were likely formed along parallel edges of neighboring bilayer islands, reflecting the gap between the bottom layers of the bilayer islands, whereas the top layers of the bilayer islands were continuously connected to each other.

The atomically resolved STM images (Fig. 2c and Fig. S6a) show that the lattice constant of the $MoS_2$ monolayer is 3.19 ± 0.05 Å and the lattice constant of the top layer ($WS_2$) of the heterobilayer is 3.20 ± 0.02 Å, respectively. The measured lattice constants are consistent with the previous theoretical calculations and experiment results[34-38]. The fast Fourier transform (FFT) images (the insets in Fig. 2c) reveal that the same crystalline orientation of the monolayer $MoS_2$ is the same as that of the top layer of the heterobilayer, corroborating the continuity of the monolayer and the top layer of the heterobilayer (Fig. 2d). Synthesized heterostructures have wide range of applications in electronics, optoelectronics, photovoltaics and catalysis. Clarifying the interface between the $MoS_2$ monolayer and the $WS_2/MoS_2$ heterobilayers is crucial for understanding the electronic and optical properties of such devices based on these heterostructures.

The height difference between the heterobilayer and the monolayer can be determined from the height histogram of an area straddling across the interface (the larger rectangle in Fig. 2b). The height histogram was fit well to a two-peak Gaussian function, reflecting a clear Gaussian distribution of heights for both the monolayer and heterobilayer (Fig. 3a). The only deviation from the fitting occurs around the middle of the two peaks, which is attributed to the continuous transition between the $MoS_2$ monolayer and the heterobilayer. For comparison, the height histogram of the areas away from the interfaces (areas marked in Fig. S6c) was fit perfectly to a two-peak Gaussian distribution with a zero probability in the middle of the two peaks (Fig. 3b). The height difference between the heterobilayer and the monolayer is determined by the distance between the centers of the two peaks to be 0.837 ± 0.085 nm. This measured height is consistent with the height of a $MoS_2$ or $WS_2$ monolayer in the previous reports[14, 30, 35]. From the width of the Gaussian peaks in the height histograms, we can also determine the roughness of $MoS_2$ monolayer and heterobilayer on $SiO_2$ surfaces. To reduce estimation errors due to the continuous interfaces between the monolayer and heterobilayer, we averaged the roughness evaluated from several areas on the monolayer and the heterobilayer away from the interfaces. The average RMS roughness is 0.329 ± 0.033 nm for the $MoS_2$ monolayer region and 0.230 ± 0.021 nm for the heterobilayer. Our result is consistent with the previously reported roughness of a $MoS_2$ monolayer on $SiO_2$[36]. The monolayer roughness is also compatible with the roughness of a bare $SiO_2$ substrate previously measured by AFM[39-40], indicating that the $MoS_2$ monolayer conformed well to the underneath $SiO_2$ surface. The slightly smoother topography of the heterobilayer is expected since the bending stiffness of the 2D material increases with thickness. A similar trend was observed in bilayer and monolayer graphene on $SiO_2$[41]. Moreover, the roughness of the heterobilayer is compatible with the previously reposted roughness of a $MoS_2$ bilayer on $SiO_2$[36], implying that the stiffness of the heterobilayer is similar to that of $MoS_2$ bilayer.

After identifying the structures of the heterobilayer, the $MoS_2$ monolayer and their interface, now we turn to the electronic properties of these structures. We performed STS measurements at room temperature. At 77 K, the conductivity of the samples was too low to measure tunneling currents and the STM tip would crash into the samples. Therefore, all the STM/STS measurements were performed at room temperature. Fig. 4a shows the representative dI/dV curves obtained at the $MoS_2$ monolayer and the heterobilayer. In the averaged dI/dV curve, we linearly fitted the sloped parts that correspond to the conduction band. The conduction band edges are determined by extrapolating these linear lines to their crossings with the minimum conductance, while the valence band edges are determined by the peak positions at negative bias. The band gap is estimated by the separation of the valence band edge and conduction band edge (as shown in Fig. S7). For the $MoS_2$ monolayer, the measured band gap is 2.05 eV, with a conduction band edge at 0.75 eV and a valance band edge at −1.3 eV. These measured values are close to previously reported results on exfoliated monolayer $MoS_2$ on $SiO_2$[18, 28]. For the heterobilayer, the band gap extracted from the averaged dI/dV curve is 1.35 eV, consistent with the previous results on $MoS_2$/$WS_2$ vertical heterobilayer on $SiO_2$ fabricated via the exfoliation method[18]. At the interface region, we performed the STS measurements at different locations traversing the continuous interfaces (Figs. 4c and 4d). The transition from the heterobilayer type to the MoS$_2$ monolayer type occurs between dI/dV curves P8 and P9 in Figs. 4c and 4d. Although the electronic states previously found in open edges exhibited a distinct edge state[28], our results demonstrate the absence of edge states over the continuous transition regions between the MoS$_2$ monolayer and the heterobilayer.

To confirm our STS results, we carried out DFT calculations on a WS$_2$/MoS$_2$ heterobilayer, as well as individual MoS$_2$ and WS$_2$ monolayers (see Methods). In the simulation, the height of the heterobilayer differs from that of the monolayer by 6.15 Å, which is compatible to the measured height difference by STM. The theoretical height difference is from the bulk MoS$_2$. The calculated band structures (Fig. S8) show that the MoS$_2$ monolayer has a direct band gap of 1.72 eV at the K point, whereas the WS$_2$/MoS$_2$ heterobilayer has an indirect band gap of 1.26 eV with the valance band maximum (VBM) at the $\Gamma$ point and the conduction band minimum (CBM) located between the K and $\Gamma$ points. The decrease in band gap in the heterobilayer is in qualitatively good agreement with the experimental data, considering that DFT typically underestimates a band gap for semiconductors. As shown in Fig. 4b, when the Fermi level is set to zero, our calculated DOS plots for the MoS$_2$ monolayer and for the upper layer of the heterobilayer reveal a similar trend to the experimental dI/dV spectra below the Fermi level.

The DOS projected onto the top WS$_2$ layer in the heterobilayer (Fig. 4b) shows distinctive features near the Fermi level, compared to a free standing WS$_2$ monolayer (Fig. S9b). This change in the DOS may be attributed to the interlayer coupling between the MoS$_2$ and WS$_2$ monolayers, as explained below. To further analyze the nature of such coupling between the two different monolayers, as shown in Fig. S10, we calculated site- and orbital projected density of states (PDOS) for each atom in the MoS$_2$ and WS$_2$ monolayers and the heterobilayer, and clarified orbital characters of the VBM and CBM in each system. In the monolayers (Figs. S10i and S10j, where the WS$_2$ case is not shown since it is similar to the MoS$_2$ case), a contribution to the VBM arises mainly from the $d_{x2-y2}$ and $d_{xy}$ orbitals of the transition metal atoms with a small contribution from the $p_x$ and $p_y$ orbitals of the sulfur atoms. The $x$, $y$, and $z$ coordinates are illustrated in Fig. S10K. The CBM arises mainly from the $d_{z2}$ orbital of the metal atoms and the $p_x$ and $p_y$ orbitals of the sulfur atoms. Our results in the orbital analysis are consistent with the previous DFT study[37]. Now regarding the heterobilayer (Figs. S10c-h), we find that the states near the CBM predominantly originate mainly from the bottom MoS$_2$ layer, which corroborates with the type II band offset in the heterobilayer. The characteristics of the dominant orbitals near the CBM are not modified compared to the monolayer cases, which indicates that there is no effect of the interlayer coupling to the states near the CBM for the heterobilayer. This agrees with the experimental finding that the region near the CBM has not been affected upon building the heterobilayer (Fig. 4a). However, the orbital contributions near the VBM of the heterobilayer qualitatively change from the monolayer cases. In the heterobilayer, the DOS near the VBM mainly arises from the $d_{z2}$ orbitals of the molybdenum and tungsten atoms and the $p_z$ orbitals of inner sulfur atoms, S$_{Mo,in}$ and S$_{W,in}$, as indicated in Fig. S10. Outer sulfur atoms (S$_{Mo,out}$ and S$_{W,out}$ in Fig. S10b) have negligible contributions to the VBM. The slightly higher contribution of the $d_{z2}$ orbitals of the tungsten atom than those of the molybdenum in the close proximity to the VBM, is consistent with the type II band offset. However, the $d_{z2}$ orbitals of both the molybdenum and tungsten atoms almost equally contribute to the shoulder region below the VBM (Figs. S10c and S10f). This shoulder region for the top WS$_2$ layer in the bilayer starts to appear about 0.25 eV below that in the MoS$_2$ monolayer (Fig. 4b), which agrees with the STS data (Fig. 4a). This change of the orbital characteristics in proximity to the VBM suggests substantial hybridization between the metal $d_{z2}$ orbitals and the inner sulfur $p_z$ orbitals driven by a substantial interlayer interaction such as σ bonding.

**Conclusion**

In conclusion, we studied the morphology and electronic structures of as-grown MoS$_2$ monolayer and MoS$_2$/WS$_2$ heterobilayer on SiO$_2$ synthesized through the CVD method. The heterobilayer has a larger height than the monolayer with a height difference of 0.837 nm. The RMS roughness of the heterobilayer is 0.230 nm, compared with 0.329 nm for the monolayer. Most strikingly, we observed, for the first time, to the best of our knowledge, a closed interface between the MoS$_2$ monolayer and the heterobilayer with the atomic resolution. The discovery of such a nanostructure could deepen our understanding of the growth mechanism, interlayer interactions and electronic structures of 2D TMD heterostructures synthesized via CVD. Furthermore, our STS results and DFT calculations revealed band gaps of the heterobilayer and the MoS$_2$ monolayer agree with previously reported values for MoS$_2$ monolayer and MoS$_2$/WS$_2$ heterobilayer on SiO$_2$ fabricated through the mechanical exfoliation method. Our results could induce high impacts on optimizing and designing new 2D heterostructures.

**Methods**

**Synthesis of MoS$_2$/WS$_2$ heterobilayers**

Sulfur pieces (Alfa Aesar, 99.999 %, 1.5 g) and WO$_{3-x}$/MoO$_{3-x}$ core/shell nanowires on ~2 mm×3 mm carbon fabric strips were used as S, W and Mo precursors, respectively[21]. The carbon fabric strip with WO$_{3-x}$/MoO$_{3-x}$ core-shelled nanowires was directly placed on the top of a 300 nm SiO$_2$/Si substrate, and a ceramic boat with sulfur powder was placed upstream (Fig. S1). After purging the system with Ar for 15 min, the furnace was firstly heated up to 660 °C at a rate of 30 °C min$^{-1}$ with 150 sccm Ar and held for 25 min to vaporize the excessive MoO$_{3-x}$. The furnace was then heated up to 860 °C with a rate of 20 °C min$^{-1}$. When the furnace temperature reached 820°C, sulfur was heated by a heating belt with an individual temperature controller at ~200 °C. Then the furnace was cooled down naturally after maintaining 860 °C for another 30 min. The heating belt for sulfur was removed when the furnace was cooled down to 350 °C.

**Oxidative etching of the heterobilayers**

WS$_2$/MoS$_2$ heterobilayers were placed at the center of CVD furnace of a standard 1-inch quartz tube. After pumping down the system for 10 min, the furnace was heated up to 320 °C at a rate of 20 °C min$^{-1}$ and stayed for another 90 min.

**Characterization of as-grown and etched heterobilayers**

The optical images were taken with Olympus BX 51M microscope. AFM images were captured with Bruker Dimension Icon in tapping mode. Low-frequency Raman spectra were collected with WITEC RSA300+ Raman system under the 532 nm laser excitation with a power of 160 μW. Ordinary Raman spectra were carried out with Horiba-Jobin-Yvon Raman system under 532 nm laser excitation with a power of 2 mW. Raman and PL mapping images were performed by a step of 0.8 μm. The Si peak at 520 cm$^{-1}$ was used for calibration in the experiments.

**STM and STS measurements**

Au/Ti (80/10 nm) electrodes on the samples were fabricated by the masked thermal evaporation through a shadow mask. STM and STS characterizations were carried out in an ultra-high vacuum (UHV) STM system (a customized Omicron LT STM/Q-plus AFM system). The samples were annealed at 220 °C for 2 hours in the preparation chamber with a base pressure of < $10^{-10}$ mbar STM/STS measurements were then performed in the STM chamber that is connected to the preparation chamber. An optical microscope (Infinity K2) mounted to one of the optical windows of the STM chamber was used to precisely locate the STM tip to desired areas on the samples.

**DFT calculations**

We simulated $MoS_2$ and $WS_2$ monolayers and a $WS_2/MoS_2$ heterobilayers (i.e., a $WS_2$ monolayer on a $MoS_2$ monolayer) by DFT calculations using VASP[42-43]. We employed the Perdew-Burke-Ernzerhof (PBE) generalized-gradient approximation (GGA)[44] for the exchange-correlation functional and the projector-augmented wave (PAW) pseudopotentials[45]. Spin-orbit coupling was included self-consistently in all calculations. Previous first-principles studies[37, 46] reported that electronic structures of the TMD monolayers and homobilayers obtained from the PBE-GGA functional are similar to those from the GW method[47] or hybrid functionals such as Heyd-Scuseria-Ernzerhof (HSE06)[48], except for an underestimated band gap. Therefore, our choice of the functional was justified. For the $MoS_2$ and $WS_2$ monolayers, we used hexagonal unit cells with honeycomb structure (1H-$MoS_2$, 1H-$WS_2$) with experimental lattice constants and atomic coordinates[49-52]. For the heterobilayers, we assumed AA' stacking equivalent to the 2H stacking of bulk $MoS_2$ or $WS_2$[46], because previous GW and DFT calculations on TMD homolayers revealed that the AA' stacking has the lowest energy; the AB stacking has an energy about 5 meV higher energy than the AA' stacking; the two stackings show very similar electronic structure[46]. Regarding the $WS_2/MoS_2$ heterobilayer, we considered the experimental lattice constants of bulk $MoS_2$ ($WS_2$) for the $MoS_2$ ($WS_2$) monolayer, while adopting the vertical distance between monolayers, $d$, as the experimental value in bulk $MoS_2$. No further relaxation was carried out for the constructed heterobilayers. The choice of the heterobilayer geometry can be justified, considering that the lattice constants of $MoS_2$ and $WS_2$ differ by less than 1%[49-52], and that the interlayer distance *d* agrees with the corresponding value in bulk TMD, even upon geometry relaxation[46]. An energy cutoff was set to 300 eV and a *k*-point sampling of 27 × 27 × 1 was applied. A vacuum layer thicker than 20 Å was included in a unit cell to avoid artificial interactions between neighboring bilayers. Van der Waals interactions were not included since our calculated DOS and PDOS did not change with their inclusion. We obtained the vacuum energy by calculating Hartree potential along the vertical axis, averaged over in-plane directions. The saturated value of Hartree potential in the vacuum region was identified to be the vacuum energy. To compare the band extrema of different systems and get the band alignment, we used the vacuum energy of each systems as a reference energy, and set it to zero.

We calculated the dipole moment of the heterobilayer and found a small value of 2.4 x $10^{-3}$ ·Å and indeed, the change in the calculated vacuum energy was small upon including the dipole correction. Thus, we did not include the dipole correction in our calculations of the total and projected density of states, and the band structures. The interlayer distance was set to the experimental bulk value.

**Supporting information Available:**

Detailed description of the synthesis method, details about Raman and photoluminescence mapping of MoS$_2$ monolayer and WS$_2$/MoS$_2$ heterobilayer, additional STM results at various scales, and details about DFT calculations.

**Acknowledgements**

F.Z., H.Z. and C.T. acknowledge the financial support provided for this work by the U.S. Army Research Office under the grant W911NF-15-1-0414. L.J. acknowledges National Natural Science Foundation of China (No.51372134, No.21573125) and Tsinghua University Initiative Scientific Research Program. L.X. acknowledges NSFC (Nos. 21373066 and 21673058), Key Research Program of Frontier Sciences of CAS (QYZDB-SSW-SYS031), Strategic Priority Research Program of CAS (XDA09040300), Beijing Nova Program (Z151100000315081) and Beijing Talents Fund (2015000021223ZK17). Y. C. was supported by the ICTAS fellowship at Virginia Tech, and the computational support was provided by San Diego Supercomputer Center (SDSC) under DMR060009N and Virginia Tech Advanced Research Center.

**Figures**

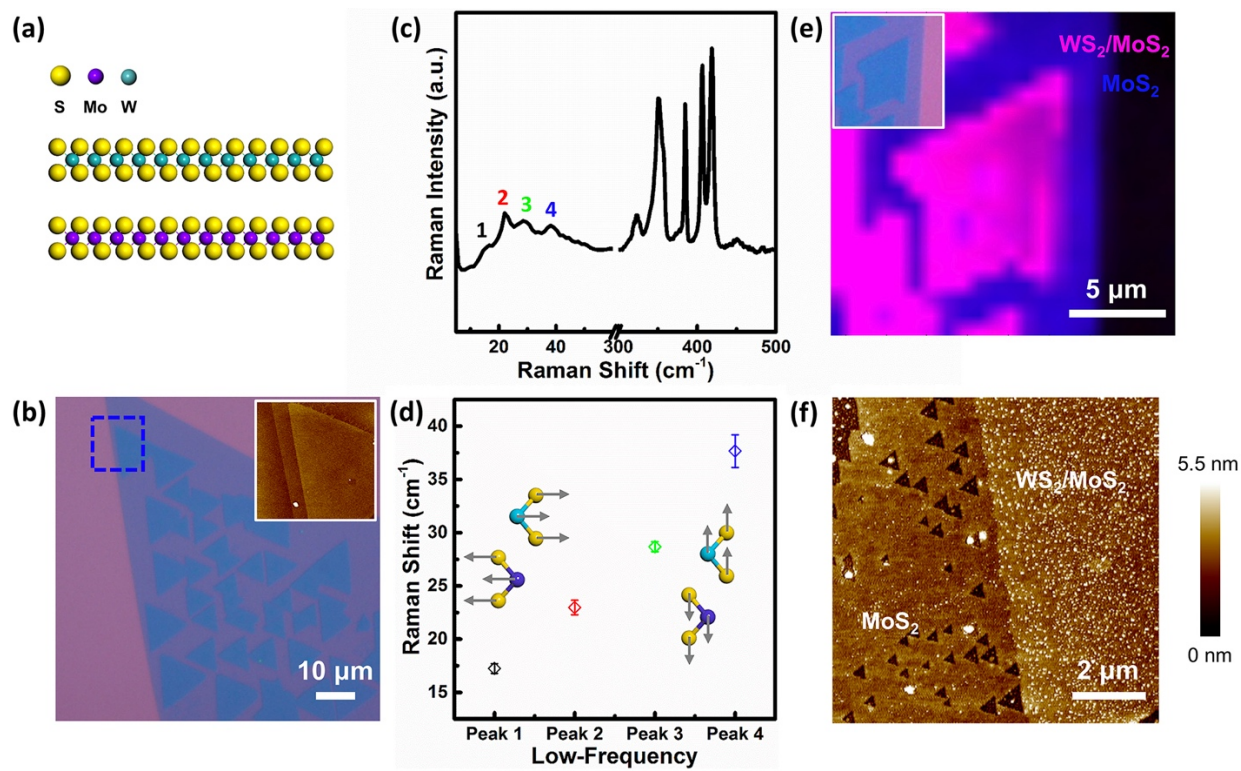

**Fig. 1.** (a) Schematic of WS$_2$/MoS$_2$ vertical heterostructures. (b) Optical image of WS$_2$/MoS$_2$ vertical heterostructures. Inset, AFM image of WS$_2$/MoS$_2$ vertical heterostructures in the blue rectangle. (c) Raman spectra of WS$_2$/MoS$_2$ vertical heterostructures. (d) Distributions of low-frequency Raman modes. Peaks 1 and 2 correspond to the shear modes, and Peaks 3 and 4 correspond to the layer breathing modes. (e) Raman mapping of MoS$_2$ monolayer and MoS$_2$/WS$_2$ vertical heterostructures in the inset. (f) AFM height image of MoS$_2$ monolayer and WS$_2$/MoS$_2$ vertical heterostructures by oxidative etching in 320 °C.

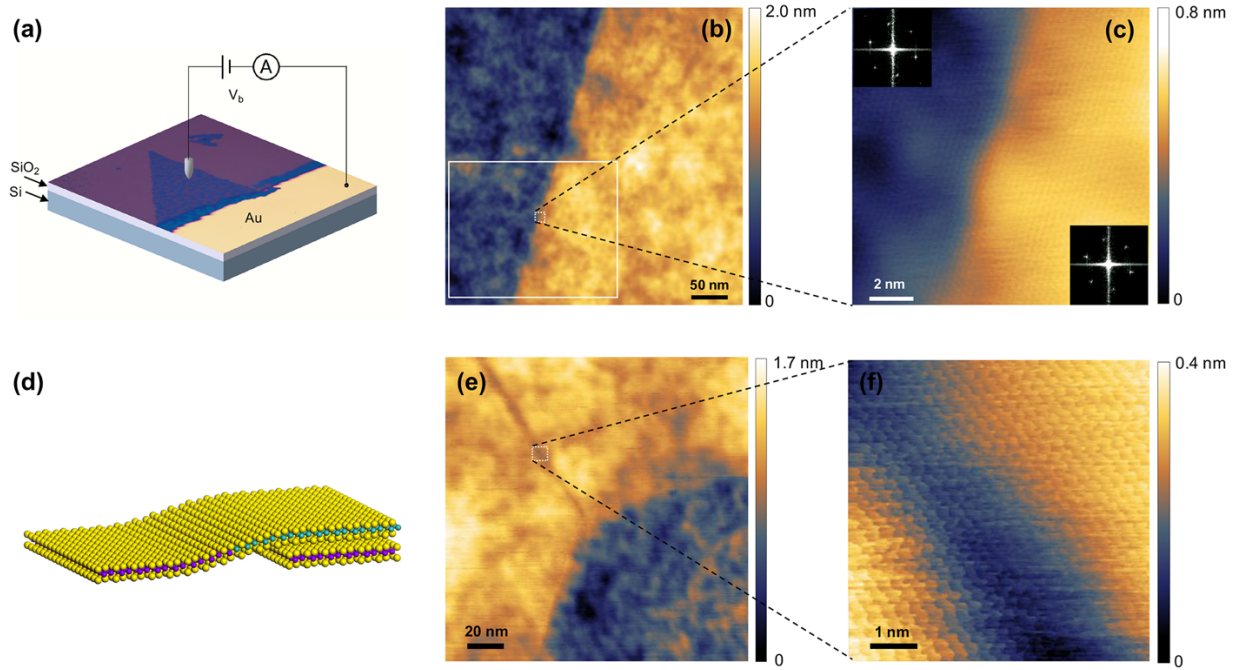

**Fig. 2.** (a) Schematic illustration of the experiment setup for STM/STS coupled with optical imaging. (b) Large-scale STM image of the MoS$_2$ monolayer and the heterobilayer ($V_s$ = 1 V, I = 0.1 nA). (c) Atomically resolved STM image of the interface between the monolayer and the heterobilayer, marked by the dashed white rectangle in (b) ($V_s$ = 1V, I = 0.15 nA). Insets, FFT of the monolayer region and heterobilayer region, respectively. (d) Schematic drawing of the interface between the heterobilayer and the MoS$_2$ monolayer. (e) Large-scale STM image of a trench structure demarcating the heterobilayer ($V_s$ = − 0.55 V, I = 0.3 nA). (f) Atomically resolved STM image of the trench structure, marked by the dashed white rectangle in (e) ($V_s$ = 1.1 V, I = 0.2 nA).

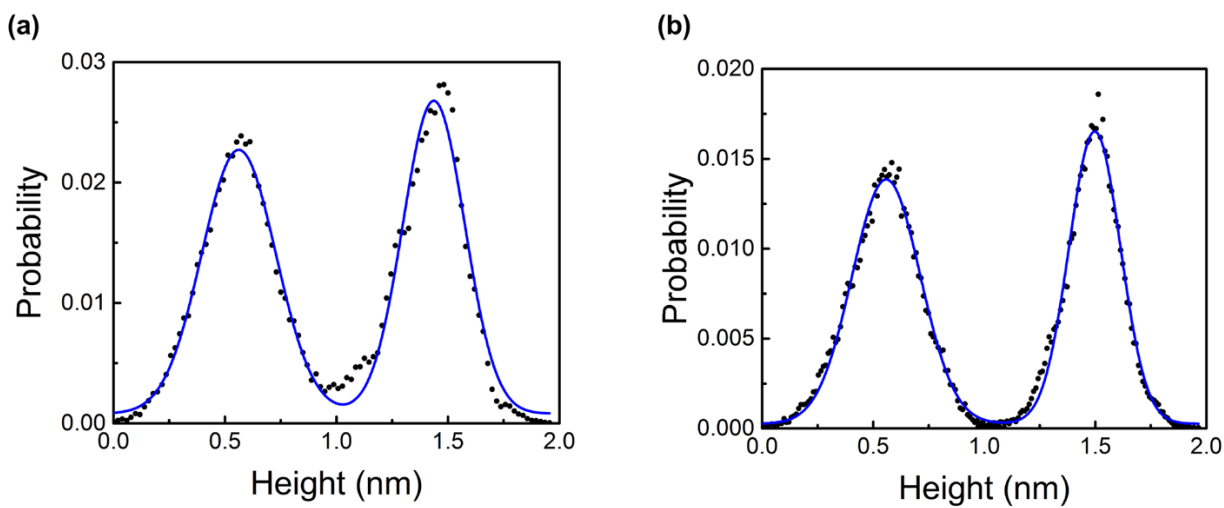

**Fig. 3.** (a) Height histogram of the area marked by the solid white rectangle in Fig. 2b, which includes the interface between the $MoS_2$ monolayer and the heterobilayer. Blue line: two-peak Gaussian fitting of the height histogram. (b) Height histogram of areas containing only the $MoS_2$ monolayer and the heterobilayer, without an interface between the $MoS_2$ monolayer and the heterobilayer. Blue line: two-peak Gaussian fitting of the height histogram.

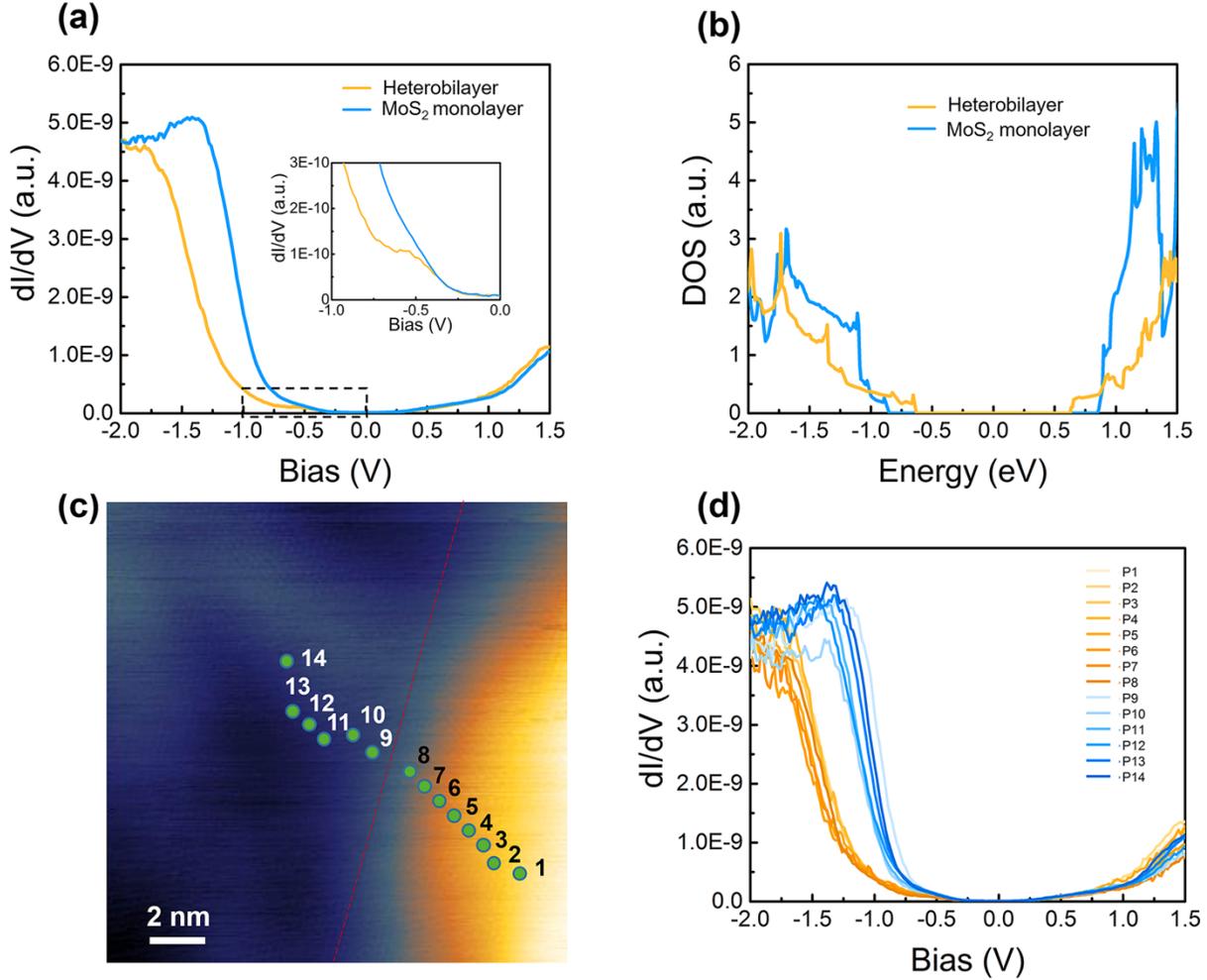

**Fig. 4.** (a) Experimental dI/dV spectra for the MoS$_2$ monolayer (blue) and the heterobilayer (yellow). Inset shows expanded view of dI/dV spectra between −1.0 and 0.0 V. (b) DFT-calculated density of states (DOS) of the MoS$_2$ monolayer (blue) and DOS projected onto the top WS$_2$ layer in the WS$_2$/MoS$_2$ heterobilayer (yellow). Here the Fermi level was set to zero. (c) STM image of the interface between the MoS$_2$ monolayer and the heterobilayer (V$_s$ = 1V, I = 0.15 nA). The interface is marked by the dashed red line. (d) dI/dV spectra acquired at the points marked in (c).